\documentclass[nofootinbib]{revtex4}
\usepackage{graphicx}
\usepackage{dcolumn}
\usepackage{amsmath,amssymb,epsfig}
\usepackage{paralist}
\usepackage{comment}
\usepackage{graphicx}
\usepackage{multirow}
\usepackage{color,soul}
\usepackage{suffix}
\usepackage{mathtools}

\renewcommand{\vec}[1]{\boldsymbol{\mathrm{#1}}}

\begin{document}

\title{Imaging point sources with the gravitational lens of an extended Sun}

\author{Slava G. Turyshev$^{1}$, Viktor T. Toth$^2$}

\affiliation{\vskip 3pt
$^1$Jet Propulsion Laboratory, California Institute of Technology,\\
4800 Oak Grove Drive, Pasadena, CA 91109-0899, USA}

\affiliation{\vskip 3pt
$^2$Ottawa, Ontario K1N 9H5, Canada}

\date{\today}

\begin{abstract}

We study the optical properties of the solar gravitational lens (SGL) while treating the Sun as an extended, axisymmetric and rotating body. The gravitational field of the Sun is represented using a set of zonal harmonics. We develop an analytical description of the intensity of light that is observed in the image plane in the strong interference region of a realistic SGL. This formalism makes it possible to model not only the point-spread function of point sources, but also actual observables, images that form in the focal plane of an imaging telescope positioned in the image plane. Perturbations of the monopole gravitational field of the Sun are dominated by the solar quadrupole moment, which results in forming an astroid caustic on the image plane. Consequently, an imaging telescope placed inside the astroid caustic observes four bright spots, forming the well-known pattern of an Einstein cross. The relative intensities and positions of these spots change as the telescope is moved in the image plane, with spots merging into bright arcs when the telescope approaches the caustic boundary. Outside the astroid caustic, only two spots remain and the observed pattern eventually becomes indistinguishable from the imaging pattern of a monopole lens at greater distances from the optical axis. We present results from extensive numerical simulations, forming the basis of our ongoing study of prospective exoplanet imaging with the SGL. These results are also applicable to describe a large class of gravitational lensing scenarios involving axisymmetric lenses that can be represented using zonal harmonics.

\end{abstract}


\maketitle

\section{Introduction}

Direct imaging of exoplanets is a challenging task.  These objects are not self-luminous, they are small and they are very far from us.  In addition, light from these sources arrives in the presence of many unwanted signals that represent various observational noises (i.e., light contamination from the host star, zodiacal light in the parent stellar system, background stars, etc.) As it is often the case in astronomy, detection sensitivity improves with the size of the telescope.  However,  there are practical limits when it comes to the size of optical instruments. For instance, direct observation of our own Earth from the distance of 100 light years, resolving it as just a single pixel, a telescope aperture of $\sim 90$~km would be required. Higher resolution requires even larger apertures, which are obviously impractical. Very long baseline optical interferometry may offer solutions when it comes to angular resolution but does not resolve the problem of faintness: to achieve the desired light amplification, a very large light collecting area is necessary.

These challenges motivate us to consider other physically permissible ways to construct an imaging instrument with superior sensitivity and resolving power.  We study the solar gravitational lens (SGL) as the means to achieve this objective \cite{Turyshev-Toth:2017}. The SGL results from the gravitational bending of light by the gravitational field of the Sun.  The angle by which a ray of light is deflected gravitationally is $\alpha=2r_g/b$, where $r_g=2GM_\odot/c^2$ is the Schwarzschild radius of the Sun and $b$ is the ray's impact parameter. Obviously, $b$ cannot be smaller than the solar radius $R_\odot$, as rays with impact parameters $b\lesssim R_\odot$ are obscured by the solar disk. Consequently, the bending-of-light effect of the Sun is rather weak, placing the focal region of the SGL (where, depending on the impact parameter, light rays from a distant source intersect) beyond $z=R_\odot^2/(2r_g)=547.8$ astronomical units (AU) from the Sun. Positioning a spacecraft with a modest telescope in that region, we benefit from the very large light amplification of the SGL, given as $\mu_0=2\pi k r_g=1.17\times 10^{11}\, (1\,\mu{\rm m}/\lambda)$, where $k=2\pi/\lambda$ is the wavenumber of an electromagnetic (EM) wave corresponding to its wavelength $\lambda$.

Modeling the SGL requires knowledge of its point-spread function (PSF). The PSF characterizes the impulse response of the optical system: it maps light from a point source into the image plane. Our investigations began with treating the Sun as a point source of a monopole gravitational field, leading to a PSF that is known very well (see \cite{Turyshev-Toth:2017} and references therein). We also studied image formation by the monopole gravitational lens \cite{Turyshev-Toth:2020-extend,Toth-Turyshev:2020}. However, the actual, physical Sun is not a perfectly structureless object with unbroken spherical symmetry. It rotates and consequently, it is slightly oblate \cite{Turyshev-Toth:2021-multipoles}. Thus, light rays propagating from  different directions along the solar limb are affected slightly differently by solar gravity and thus they are deflected by slightly different angles. As a result, as the rays continue to travel along their paths along these slightly different directions, they reach the focal region of the SGL at different points. Instead of converging on the optical axis, some rays of light miss the optical axis altogether, intersecting the image plane elsewhere. In certain regions of the image plane, this can yield strong constructive interference resulting in the formation of caustics in the image plane \cite{Turyshev-Toth:2021-multipoles,Turyshev-Toth:2021-caustics}. Elsewhere, destructive interference may produce new patterns of alternating bright and dark regions, overlapping with, and altering, the concentric pattern that is characteristic of the monopole lens. Correct representation of this behavior requires a wave-optical treatment of the gravitational lensing phenomena.

Recently, we developed such a wave-theoretical treatment for light propagation in a gravitational field of an extended, axisymmetric and rotating body with its gravity field characterized by a complete set of zonal harmonics \cite{Turyshev-Toth:2021-multipoles} perturbing the monopole gravitational field. Based on that approach, we were able to describe gravitational lensing in the presence of an extended axisymmetric gravitational lens such as our Sun. Here, we  continue to study the optical properties of the SGL using this new, much improved treatment of gravitational lensing. Our current objective is to study the images formed by point sources of light in the presence of the extended Sun, especially as it may be applicable for prospective use of the SGL for direct imaging and spectroscopy of exoplanets.

This paper is organized as follows:
In Section \ref{sec:imaging} we discuss our new angular eikonal method, which yields a new diffraction integral that describes the light diffraction in the strong interference region of an extended gravitational lens. The external gravitational field is characterized by using a gravitational monopole that is perturbed by an infinite set of zonal harmonics. We discuss imaging of point sources with such an SGL by convolving the PSF of the extended Sun with that of an imaging telescope, modeled in our case as a simple thin lens. In Section \ref{sec:numerical}, we numerically evaluate diffraction integrals involved in characterizing the optical properties of the extended axisymmetric lens. We discuss the resulting views seen by the imaging telescope at various locations in the image plane using a set of images that are also assembled in the form of short animations, presented as supplementary material. Lastly, in Section~\ref{sec:end} we discuss results and outline the next steps in our investigation.

\section{Imaging with the SGL}
\label{sec:imaging}

We characterize the SGL using its PSF. On its own, the PSF of a gravitational lens characterizes the distribution of light in an image plane. For practical imaging scenarios, this is not sufficient. Light intensity in the image plane is sampled by an optical instrument, such as a telescope, that ``looks back'' at the lens (see relevant geometry in Fig. 4 of \cite{Turyshev-Toth:2020-extend}). Such an instrument observes the lens (i.e., the solar disk) surrounded by a full Einstein ring or partial spots and arcs, as determined by the multipole moments characterizing the lens and the shape and extent of the distant light source. This combination of the SGL with an optical instrument is achieved by convolving the SGL PSF with that of the instrument.

\subsection{Optical properties of the SGL}
\label{sec:opt-prop}

The presence of gravitational multipole moments changes the conditions for the diffraction of light in the gravitational field \cite{Turyshev-Toth:2021-multipoles}. We characterize an EM wave with its wavenumber $k$, and the SGL by the Sun's Schwarzschild radius $r_g$. For a high-frequency plane EM wave (i.e., neglecting terms $\propto(kr)^{-1}$) and for $r\gg r_g$, we derive the field near the optical axis at heliocentric distance $z$, up to terms of ${\cal O}(\rho^2/z^2)$, in the form
{}
\begin{eqnarray}
    \left( \begin{aligned}
{E}_\rho& \\
{H}_\rho& \\
  \end{aligned} \right) =\left( \begin{aligned}
{H}_\phi& \\
-{E}_\phi& \\
  \end{aligned} \right) &=&
E_0
 \sqrt{2\pi kr_g}e^{i\sigma_0}B(\vec x)
  e^{i(kz -\omega t)}
 \left( \begin{aligned}
 \cos\phi& \\
 \sin\phi& \\
  \end{aligned} \right),
  \label{eq:DB-sol-rho}
\end{eqnarray}
with the remaining components being small, $({E}_z, {H}_z)= {\cal O}({\rho}/{z})$, or constant, $\sigma_0=-kr_g\ln kr_g/e-{\textstyle\frac{\pi}{4}}$ \cite{Turyshev-Toth:2019}. In this formulation, we use the following notations for the vector of the impact parameter, $\vec b$, coordinates on the image plane, $\vec x$, and the unit vector in the direction of the solar axis of rotation, $\vec s$:
\begin{eqnarray}
{\vec b}&=&b(\cos\phi_\xi,\sin \phi_\xi,0), \\
{\vec x}&=&\rho(\cos\phi,\sin \phi,0),\\
{\vec s}&=&(\sin\beta_s\cos\phi_s,\sin\beta_s\sin\phi_s,\cos\beta_s).
\label{eq:note}
\end{eqnarray}

The quantity $B(\vec x) \equiv B(\rho,\phi)$,  is the complex amplitude of the EM field in the image plane,
{}
\begin{eqnarray}
B(\vec x) &=&
\frac{1}{2\pi}\int_0^{2\pi} d\phi_\xi \exp\Big[-ik\Big(\sqrt{\frac{2r_g}{r}}\rho\cos(\phi_\xi-\phi)+
2r_g\sum_{n=2}^\infty \frac{J_n}{n} \Big(\frac{R_\odot }{\sqrt{2r_gr}}\Big)^n\sin^n\beta_s\cos[n(\phi_\xi-\phi_s)]\Big)\Big],~~~~
  \label{eq:B2}
\end{eqnarray}
after it scatters on the  gravitational field of an extended lens that is characterized by zonal harmonics.

Eq.~(\ref{eq:B2}) is a new diffraction integral formula that extends previous wave-theoretical descriptions of gravitational lensing to a lens with an arbitrary axisymmetric gravitational field.  This new result, obtained through what we call the angular eikonal method \cite{Turyshev-Toth:2021-multipoles}, offers a powerful tool to study gravitational lensing in the limit of weak gravitational fields in the first post-Newtonian approximation of the general theory of relativity.

Using (\ref{eq:DB-sol-rho}), we can compute the energy flux in the image region of the lens. With overline and brackets denoting time averaging and ensemble averaging, the relevant components of the time-averaged Poynting vector for the EM field in the image volume may be given in the following form (see \cite{Turyshev-Toth:2017,Turyshev-Toth:2019,Turyshev-Toth:2020-extend} for details):
$S_z({\vec x})=(c/4\pi)\big<\overline{[{\rm Re}{\vec E}\times{\rm Re}{\vec H}]}_z\big>= (c/4\pi)E_0^2\,{2\pi kr_g}\,
\big<\overline{\big({\rm Re}\big[{ B}({\vec x})e^{i(kz-\omega t)}\big]\big)^2}\big>,$
with ${\bar S}_\rho= {\bar S}_\phi=0$ for all practical purposes.

Defining light amplification as usual \cite{Turyshev-Toth:2017,Turyshev-Toth:2019,Turyshev-Toth:2020-extend}, $\mu_z({\vec x})=S_z({\vec x})/|\vec S_0({\vec x})|$, where where $\vec S_0({\vec x})=({c}/{8\pi})E_0^2\, \vec k$ is the Poynting vector carried by a plane wave in a vacuum in flat spacetime, we have the light amplification factor of the lens that, for short wavelengths (i.e., $kr_g\gg1$) is given by
 {}
\begin{eqnarray}
\mu_z({\vec x})={2\pi kr_g} \, {\rm PSF}({\vec x}) \qquad {\rm where }\qquad\, {\rm PSF}({\vec x}) = |B({\vec x})|^2,
  \label{eq:S_mu}
   \label{eq:psf}
\end{eqnarray}
with $|B({\vec x})|^2=B({\vec x})B^*({\vec x})$, with $B^*(\vec x)$ being the complex conjugate of $B({\vec x})$ from (\ref{eq:B2}), is the PSF of the SGL.

 \begin{figure}
\raisebox{3.2in}{a)}~\includegraphics{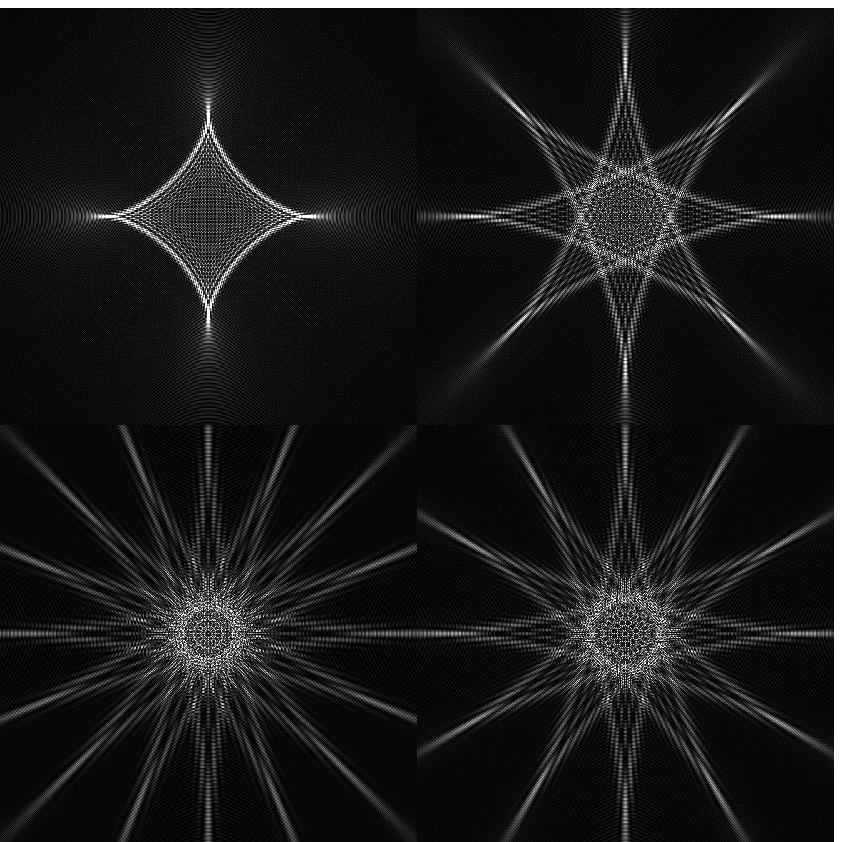}~\raisebox{3.2in}{b)}~\includegraphics{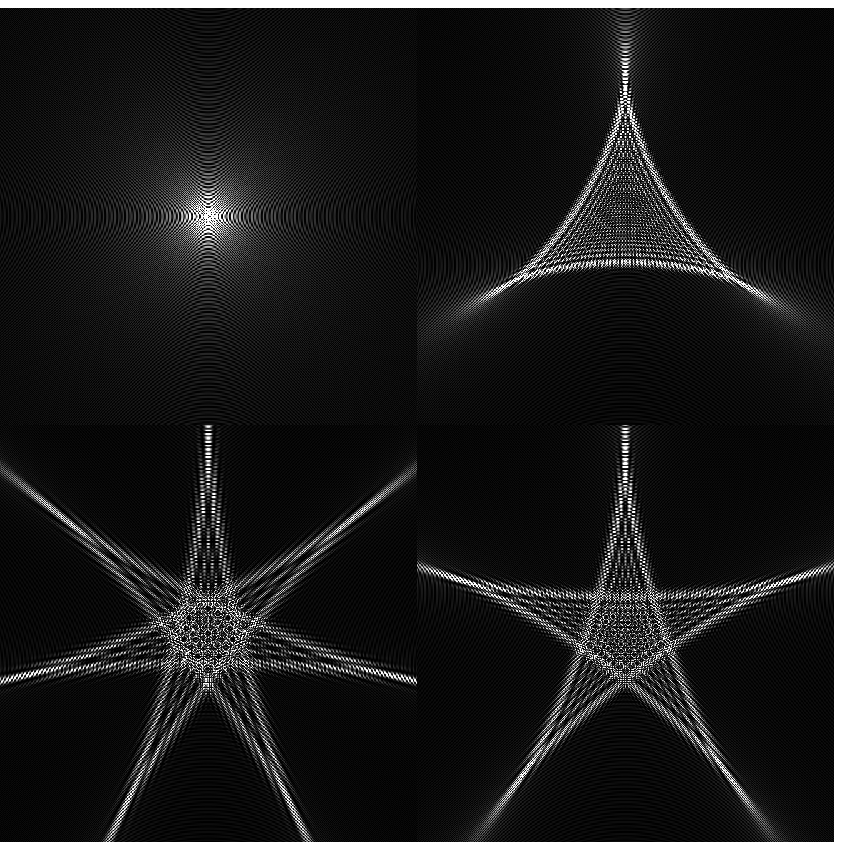}
\caption{\label{fig:caustics} Caustics representing individual contributions of the zonal harmonics of a gravitational field to the PSF (adapted from \cite{Turyshev-Toth:2021-multipoles}). Depicted are images of a point source, formed by a lens with a specific zonal harmonic, with other harmonics suppressed. The images were obtained via numerical integration of ${\rm PSF}=|B({\vec x})|^2$ with $B({\vec x})$ from (\ref{eq:B2}). From top left, clockwise: a) $J_2$, $J_4$, $J_6$, $J_8$; b) monopole, $J_3$, $J_5$, $J_7$. For the odd-numbered caustics, a change in sign flips the image in the north-south direction.
}
\end{figure}

As we see from (\ref{eq:S_mu}), light amplification by the lens is driven by the factor $\mu_0={2\pi kr_g}$. However, its overall behavior is modified. The PSF of the lens is  extended from  $J^2_0(k\sqrt{{2r_g}/{r}}\rho)$, the form established for the monopole lens (as discussed in  \cite{Turyshev-Toth:2017}) and takes the form of $|B({\vec x})|^2$ that now provides a complete description of the intensity distribution in the image plane and accounts for gravitational lensing by an arbitrary axisymmetric gravitational potential.

${\rm PSF}({\vec x})$ determines the density of the EM field in the image plane in the strong interference region of the SGL. This function governs the optical properties of the SGL as far as imaging is concerned, as it describes light received in the image plane from a point source at infinity. Fig.~\ref{fig:caustics} shows that $|B({\vec x})|^2$ reaches its maximum value on the caustic that is formed in the image plane. For lenses dominated by the contribution of a single multipole moment, these caustics acquire the shapes of hypocycloids \cite{Turyshev-Toth:2021-caustics} (e.g., the astroid, characterizing the $J_2$ quadrupole). However, when several multipole moments are present, their interaction results in more complex shapes (see \cite{Turyshev-Toth:2021-multipoles,Turyshev-Toth:2021-caustics} for discussion.)

As we apply these results to the SGL, we recognize the fact that the Sun is an axisymmetric rotating body and as such, it will have only even multipole moments, namely $J_2, J_4, J_6,J_8$, etc. determined as $J_2=(2.25\pm0.09)\times 10^{-7}$ \cite{Park-etal:2017},  and $J_4=-4.44\times 10^{-9}$, $J_6=-2.79\times 10^{-10}$, $J_8=1.48\times 10^{-11}$ \cite{Roxburgh:2001}. The $J_{10}$ and higher multipole moments have negligible effect on the SGL's diffraction pattern, thus may be omitted. With these values, we obtain the most comprehensive form of the complex amplitude of the EM field in the strong interference region of the SGL.

\subsection{Image formation by an optical telescope in the SGL image plane}
\label{sec:image-form-Fourier}

Describing the caustic that is produced by the SGL in the image plane is important, but it is just as important to realize that the caustic is not a practical observable. The caustic pattern is projected onto the image plane, where it would be visible if its light was captured by a very large surface, i.e., a projection screen. This is obviously not practical.

Instead, we expect that the SGL will be used in conjunction with an imaging telescope that looks back at the Sun. This telescope would see the Sun (likely blocked by a suitable coronagraph), surrounded by the solar corona, through which a distant source that is lined up with the telescope would appear in the form of an Einstein ring. This Einstein ring, or parts thereof, as it appears on the telescope's imaging sensor, is our primary observable.

To model this view, we represent the imaging telescope by a  convex lens with aperture $d$ and focal distance $f$ (again, see Fig. 4 of \cite{Turyshev-Toth:2020-extend}). The relevant geometry is described by several parameters, such as ${\vec x}$ being the current position of an optical telescope in the SGL's image plane, ${\vec x}'$, being any point on the same plane within the telescope aperture positioned at ${\vec x}$, and ${\vec x}_i$, being a point on the focal plane of the optical telescope. These positions are given as
{}
\begin{eqnarray}
\{{\vec x}\}&\equiv& (x,y,0)=\rho\big(\cos\phi,\sin\phi,0\big)=\rho\,{\vec n}, \label{eq:coord'}\\
\{{\vec x}'\}&\equiv& (x',y',0)=\rho'\big(\cos\phi',\sin\phi],0\big)=\rho'\,{\vec n}',
\label{eq:x-im} \\
 \{{\vec x}_i\}&\equiv& (x_i,y_i)=\rho_i\big(\cos\phi_i,\sin\phi_i\big)=\rho_i{\vec n}_i.
  \label{eq:coord}
\end{eqnarray}

As $\vec x$ is the position of the telescope in the image plane, we present $\vec x$ as $\vec x ~ \Rightarrow ~ {\vec x}+\vec x'$, where $\vec x'$ varies only with the aperture, while  $\vec x$ covers the entire image area. Therefore, we may present the monopole term in (\ref{eq:B2})  as
{}
\begin{eqnarray}
\rho\cos(\phi_\xi-\phi)=({\vec n}_\xi\cdot{\vec x}) \qquad \rightarrow \qquad
({\vec n}_\xi\cdot({\vec x}+{\vec x}'))=
\rho\cos(\phi_\xi-\phi)+\rho'\cos(\phi_\xi-\phi').
\label{eq:rn2}
\end{eqnarray}
As a result, the complex amplitude (\ref{eq:B2}) takes the form \cite{Turyshev-Toth:2021-multipoles}
{}
\begin{eqnarray}
B(\vec x,\vec x') &=&
\frac{1}{2\pi}\int_0^{2\pi} d\phi_\xi \exp\Big[-ik\Big(\sqrt{\frac{2r_g}{r}}\Big({\vec n}_\xi\cdot\big({\vec x}+{\vec x}'\big)\Big)
+2r_g\sum_{n=2}^\infty \frac{J_n}{n} \Big(\frac{R_\odot }{\sqrt{2r_gr}}\Big)^n\sin^n\beta_s\cos[n(\phi_\xi-\phi_s)]\Big)\Big].~~~~
  \label{eq:B2+}
\end{eqnarray}

The presence of a convex lens is equivalent to a Fourier transform of the wave amplitude. We position the telescope at a point with coordinates ${\vec x}'$ in the image plane in the strong interference region of the lens
\cite{Turyshev-Toth:2019-extend,Turyshev-Toth:2020-image}.  To stay within the image, ${\vec x}'$ is within the range:  $|{\vec x}'|+d/2\leq r_\oplus$. The amplitude of the EM wave just in front of the telescope aperture, from (\ref{eq:DB-sol-rho}), is given as $B({\vec x}, {\vec x}')$.

The focal plane of the optical telescope is located at the focal distance $f$ of the lens, centered on ${\vec x}'$. Using the Fresnel--Kirchhoff diffraction formula, the amplitude of the image field in the optical telescope's focal plane at a location ${\vec x}_i=(x_i,y_i)$ is given by  \cite{Wolf-Gabor:1959,Richards-Wolf:1959,Born-Wolf:1999}:
{}
\begin{eqnarray}
{B}({\vec x},{\vec x}_i)=\frac{i}{\lambda}\iint \displaylimits_{|{\vec x}'|^2\leq (d/2)^2} \hskip -7pt  B({\vec x},{\vec x}')e^{-i\frac{k}{2f}|{\vec x}'|^2}\frac{e^{iks'}}{s'}d^2{\vec x}'.
  \label{eq:amp-w-f0}
\end{eqnarray}
The function $\exp[-i\frac{k}{2f}|{\vec x}'|^2]=\exp[-i\frac{k}{2f}(x'^2+y'^2)]$ represents the action of the convex lens that transforms incident plane waves to spherical waves, focusing at the focal point. Assuming that the focal length is sufficiently greater than the radius of the lens, we may approximate the optical path $s'$ as $s'=\sqrt{(x'-x_i)^2+(y'-y_i)^2+f^2}\sim f+\big((x'-x_i)^2+(y'-y_i)^2\big)/2f$. Furthermore, Eq.~(\ref{eq:B2+}) allows us to consider imaging of point sources with the SGL, now treated as that produced by a gravitating body that is axisymmetric and rotating, thus admitting characterization of its external gravitational field by zonal harmonics. To accomplish this, following \cite{Turyshev-Toth:2020-image,Turyshev-Toth:2020-extend}, we use the expression for $B({\vec x},{\vec x}')$ from (\ref{eq:B2+}) and present the Fresnel--Kirchhoff diffraction formula (\ref{eq:amp-w-f0}) as
{}
\begin{eqnarray}
{B}({\vec x},{\vec x}_i)&=& - \frac{e^{ikf(1+{{\vec x}_i^2}/{2f^2})}}{i\lambda f} \frac{1}{{2\pi}}\int_0^{2\pi} d\phi_\xi\int_0^{d/2}\hskip -12pt \rho' d\rho'  \int_0^{2\pi}  \hskip -10pt d\phi' \,\,
 \times \nonumber\\
&&
\hskip 20pt
\times \exp\Big[-ik\Big(\sqrt{\frac{2r_g}{r}}\rho'\cos(\phi_\xi-\phi')+
\frac{\rho_i}{f}\rho'\cos(\phi_i-\phi')\Big)-\nonumber\\
&&\hskip 60pt -ik\Big(\sqrt{\frac{2r_g}{r}}\rho\cos(\phi_\xi-\phi)+\,
2r_g\sum_{n=2}^\infty \frac{J_n}{n}\Big(\frac{R_\odot }{\sqrt{2r_gr}}\Big)^n\sin^n\beta_s\cos[n(\phi_\xi-\phi_s)]\Big)\Big].
  \label{eq:amp-w-f}
\end{eqnarray}

Following \cite{Turyshev-Toth:2020-extend}, we define the spatial frequencies
{}
\begin{eqnarray}
\alpha&=&k\sqrt\frac{2r_g}{r},~~~~ \eta_i=k\frac{\rho_i}{f},
\label{eq:zerJ+}
\end{eqnarray}
and transform the $\rho'$-dependent part  of the phase in (\ref{eq:amp-w-f})  as
{}
\begin{eqnarray}
\alpha\rho'\cos(\phi_\xi-\phi')+\eta_i\rho'\cos(\phi_i-\phi')=\rho'\,
u(\phi_\xi,\phi_i) \cos\big(\phi'-\epsilon\big),
  \label{eq:ph4}
\end{eqnarray}
with the quantities $u$ and $\epsilon$ are given by the following relationships:
{}
\begin{eqnarray}
u(\phi_\xi,\phi_i)=\sqrt{\alpha^2+2\alpha\eta_i\cos\big(\phi_\xi-\phi_i\big)+\eta_i^2},
\qquad
\cos\epsilon=\frac{\alpha  \cos\phi_\xi+\eta_i\cos\phi_i}{u},
\qquad
\sin\epsilon=\frac{\alpha  \sin\phi_\xi+\eta_i\sin\phi_i}{u}.
  \label{eq:eps}
\end{eqnarray}

This transforms expression  (\ref{eq:amp-w-f}) into
{}
\begin{eqnarray}
{B}({\vec x},{\vec x}_i)&=&ie^{ikf(1+{{\vec x}_i^2}/{2f^2})}
\Big(\frac{kd^2}{8f}\Big) \,
{\cal A}({\vec x},{\vec x}_i),
  \label{eq:BinscER}
\end{eqnarray}
where ${\cal A}({\vec x},{\vec x}_i)$ is given by
{}
\begin{eqnarray}
{\cal A}({\vec x},{\vec x}_i)&=&
\frac{1}{2\pi}\int_0^{2\pi} d\phi_\xi \,
  \Big(\frac{
2J_1(u(\phi_\xi,\phi_i)\frac{1}{2}d)}{u(\phi_\xi,\phi_i) \frac{1}{2}d}\Big)\times\nonumber\\
&&\times \,
\exp\Big[-ik\Big(\sqrt{\frac{2r_g}{ r}} \rho\cos(\phi_\xi-\phi)+
2r_g\sum_{n=2}^\infty  \frac{J_n}{n}\Big(\frac{R}{\sqrt{2r_g  r}} \Big)^n \sin^n\beta_s\cos[n(\phi_\xi-\phi_s)]\Big)\Big].~~~~~
  \label{eq:AinscER}
\end{eqnarray}

Similarly to (\ref{eq:S_mu}),  we obtain the light amplification factor, $\mu({\vec x},{\vec x}_i)=S_z({\vec x},{\vec x}_i)/|{\vec S}_0({\vec x},{\vec x}_i)|$  of the optical system consisting of the SGL and an imaging telescope (i.e., the convolution of the PSF of the SGL with that of an optical telescope), that in the case of the extended SGL takes the form
 {}
\begin{eqnarray}
\mu({\vec x},{\vec x}_i)={2\pi kr_g} I({\vec x},{\vec x}_i),
  \label{eq:Pv}
 \label{eq:Pvdd}
\end{eqnarray}
where $I({\vec x},{\vec x}_i)=|{\cal A}({\vec x},{\vec x}_i)|^2$ is the
intensity distribution corresponding to the image of a point source as seen by the imaging telescope.

As a result, to compute the power received by a detector in the focal plane of an imaging telescope positioned at the SGL image plane, we need to first compute the Fourier transform of the complex amplitude of the EM field and then follow the process that is outlined above and is captured by (\ref{eq:Pv}) with ${\cal A}({\vec x},{\vec x}_i)$ from (\ref{eq:Pv}). This approach makes it possible to employ the powerful tools of  Fourier optics (e.g.,  \cite{Goodman:2017})  to develop practical applications of the SGL.

\section{Numerical studies}
\label{sec:numerical}

To analyze the imaging of  point sources using extended gravitational lenses, we conducted numerical investigations of ${\rm PSF}({\vec x})=|B({\vec x})|^2$ from (\ref{eq:psf}) and the intensity distribution in the focal plane of an imaging telescope, $I({\vec x},{\vec x}_i)=|{\cal A}({\vec x},{\vec x}_i)|^2$, given by (\ref{eq:Pv}). Both expressions allow us to study these quantities as the telescope moves in the image plane. If no zonal harmonics are present, both expressions reduce to their monopole forms studied in \cite{Turyshev-Toth:2017} and \cite{Turyshev-Toth:2020-image}, correspondingly.

\subsection{Software}

In anticipation of reuse, we developed a simple modular software framework that consists of three main components:
\begin{inparaenum}[1).]
\item code that calculates the SGL PSF;
\item code that calculates, e.g., the imaging telescope PSF to be convolved with the SGL PSF;
\item a code framework that calls these subroutines as required, implements a standardized set of parameters, and outputs results.
\end{inparaenum}
Each of these components, written in the C++ language, is a modular component that can be replaced with functional equivalents.

The primary method of computing the PSF amounts to numerically evaluating the integral (\ref{eq:B2}). The phase that appears under the integral sign can vary rapidly for large values of $\rho$ or for large values of the $J_n$ coefficients. Therefore, even in a naively implemented integrator, an adaptive integration step is necessary to ensure that the result remains accurate. In particular, numerical integration of this expression becomes computationally expensive when $\rho$ is large, $k\rho\sqrt{2r_g/r}\gg{\cal O}(100)$. For these cases, we have also considered alternative integration methods; this research will be published separately.

Eq.~(\ref{eq:B2}) or, equivalently, Eq.~(\ref{eq:psf}), represents the ``raw'' PSF of the SGL, i.e., it can be used to model images that are projected onto the image plane by the Sun. In contrast, Eq.~(\ref{eq:Pvdd}) represents a convolution of the SGL PSF with the recognizable PSF of the thin lens telescope, in the form of the $J_1$ Bessel function. The simple appearance of (\ref{eq:Pvdd}) in comparison with (\ref{eq:psf}) represents
a specific application of the Fourier convolution theorem: indeed, the PSF of any other optical system could be convolved with the SGL PSF in a similar manner. In the software implementation, we took advantage of this and implemented the thin lens telescope PSF as a separate module, anticipating the possibility that this PSF may be replaced with one that characterizes a more complex imaging system, which may also incorporate coronagraph instrument which is needed to block the light from the solar disk and that from the inner regions of the solar corona \cite{Turyshev-etal:2018,Turyshev-etal:2020}.

\begin{figure}
\includegraphics{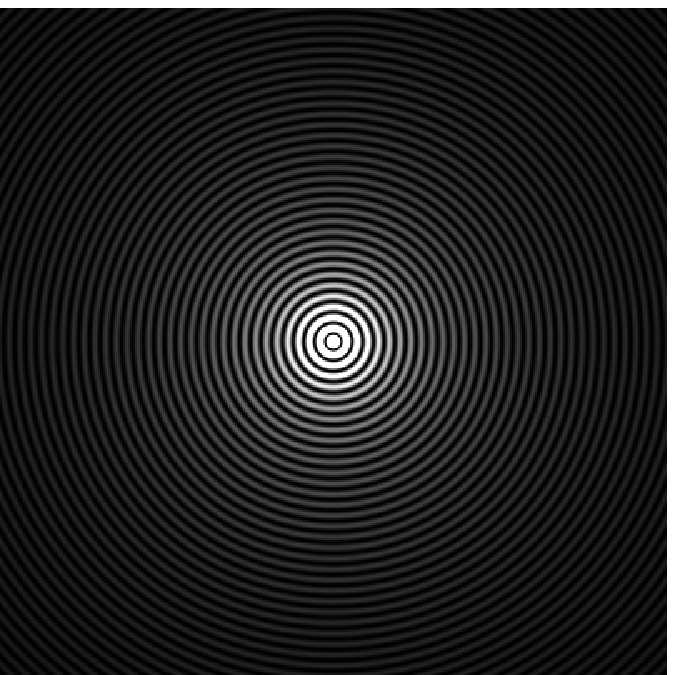}~\includegraphics{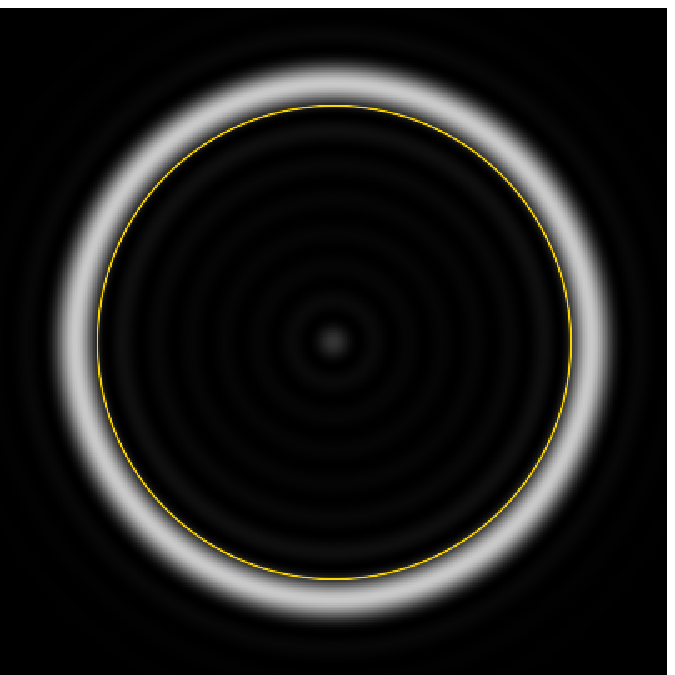}
\caption{\label{fig:monoring}The monopole PSF of the SGL and the resulting Einstein-ring as seen through an imaging telescope. The PSF (left) at $\lambda=1~\mu$m is shown in a $4\times 4~{\rm m}^2$ area at 650~AU from the Sun.The PSF consists of a concentric wave pattern. The resulting telescopic image, shown right, is as seen by a thin lens telescope with a 1~m aperture. Thin yellow circle represents the outline of the solar disk.}
\end{figure}

The software framework in which these modules are implemented provides a set of standardized parameters that can be used to control the computation. These include the origin and size of the image plane area of interest, the focal line of the optical telescope and the size of its sensors, as well as physical parameters such as the image plane distance from the Sun or the wavelength of light.

The result is a scriptable set of programs that produce binary output in the form of an array of floating point numbers, representing the intensity of light either in the image plane or on the sensor of the imaging telescope. The implementation, though not yet optimized, is nonetheless efficient enough to generate images in the matter of minutes, making it possible to produce short animations, ``movies'' using desktop-class computer hardware.

In the remainder of this section, we present some results from this simulation and offer our interpretation and insight.

\subsection{The monopole PSF}

\begin{figure}
\includegraphics{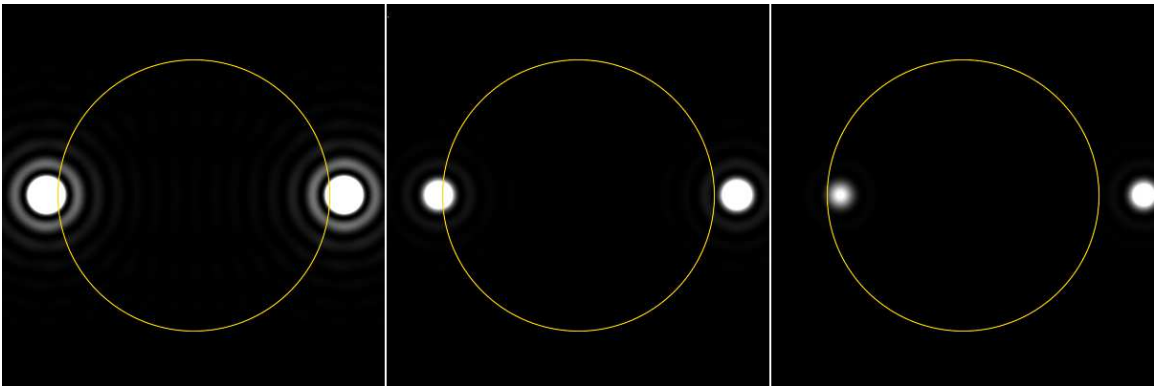}
\caption{\label{fig:approach}View of a point source (same as in Fig.~\ref{fig:monoring}) by an optical telescope with a 1~m aperture, from various vantage points in an image plane 650~AU from the Sun (indicated by the yellow circle, always centered in the image). Left: The telescope is located at $\rho=20,000$~km to the right of the optical axis; the image is still apparently symmetric, with significant light amplification evident. Middle: By $\rho=100,000$~km, significant asymmetry develops, with the ``minor'' image about to be obscured by the Sun. Right: At $\rho=300,000$~km, the ``minor'' image (which would now be substantially fainter) is fully obscured by the Sun, while the ``major'' image, no longer substantially amplified by lensing, is transitioning into the stand-along image of the point source unaffected by solar gravity.}
\end{figure}

As we observed in the monopole case \cite{Turyshev-Toth:2020-image}, if the telescope is positioned on the primary optical axis of the lens, it sees a perfect Einstein ring  (Fig.~\ref{fig:monoring}). As the telescope begins to move away from that position, increasing the separation from the optical axis, $\rho$, the ring will break into two arcs of even brightness.  As $\rho$ increases, the arcs transform into two spots with even brightness (Fig. \ref{fig:approach}). The telescope must move a large distance (comparable to the solar radius) away from the optical axis before the two spots begin to shift noticeably, and their brightness becomes uneven. For large $\rho$, eventually one spot becomes hidden behind the Sun, while the other appears at an ever greater distance from the Sun. Eventually, a very large distance away from the optical axis this spot becomes the unamplified, direct view of the point source. This behavior is axisymmetric, representing the inherent symmetries of the monopole case where the PSF
is axisymmetric.

Our region of interest in the present investigation remains the immediate vicinity of the optical axis. It is in this region that the effect of deviations from the monopole gravitational potential can be profound, with direct impact the SGL's image forming capability.

\subsection{The PSF of the extended Sun}

\begin{figure}
\includegraphics{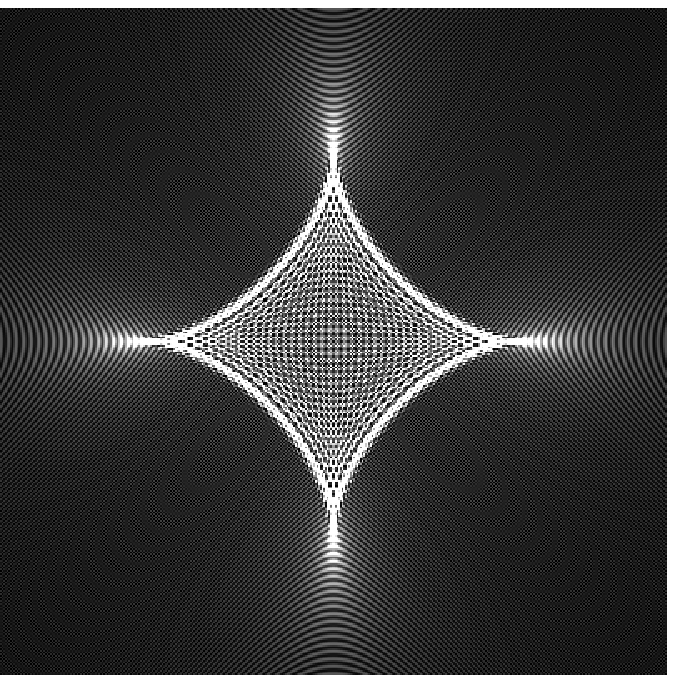}~\includegraphics{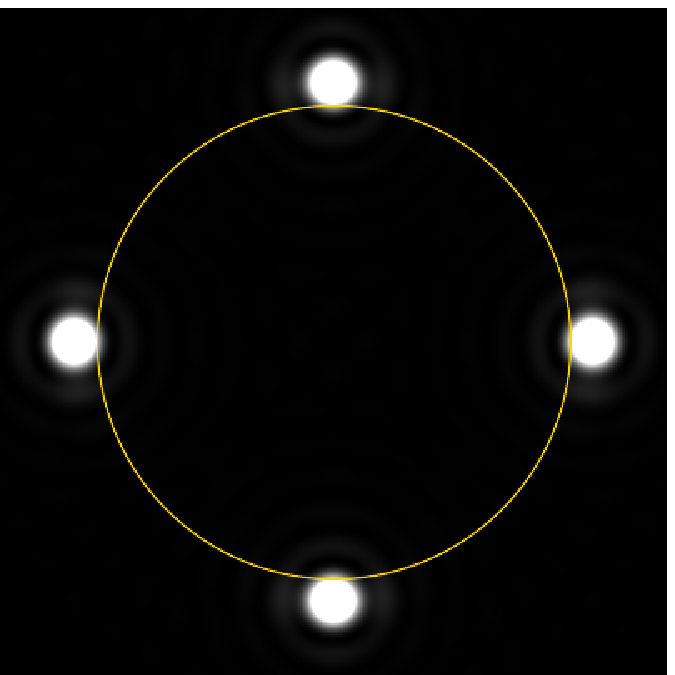}
\caption{\label{fig:psf}Left: the SGL PSF, as projected onto a $20\times 20$~m${}^2$ area in an image plane 650~AU from the Sun, at a high solar latitude, $8.13^\circ$ from the solar axis of rotation, which produces an astroid caustic that is approximately $2\rho_c^{\tt cusp}=2(4\alpha/\beta_2)\sim 11.5$~m diagonally, cusp-to-cusp. A distant point source emitting monochromatic light at $\lambda=1~\mu$m is used. Right: view from a telescope, located at the center of the PSF and aimed at the Sun. The outline of the Sun is indicated by the yellow circle.}
\end{figure}

In the case of the Sun, deviations from the monopole gravitational potential, especially at high solar latitudes, $\sin\beta_s\ll 1$, are dominated by the gravitational field's quadrupole moment, which yields the well-known astroid caustic \cite{Turyshev-Toth:2021-caustics}. An example of this is shown in Fig.~\ref{fig:psf} (left).

As we discussed in \cite{Turyshev-Toth:2021-caustics}, based on the diffraction integral (\ref{eq:B2}), and defining
{}
\begin{eqnarray}
\alpha&=&k\sqrt\frac{2r_g}{r},\qquad
\beta_2 =kr_gJ_2 \Big(\frac{R_\odot }{\sqrt{2r_gr}}\Big)^2\sin^2\beta_s,
\label{eq:kbet-J2}
\end{eqnarray}
the  astroid caustic may be given in a parametric form as
{}
\begin{eqnarray}
x&=&\frac{4\beta_2}{\alpha}\cos^3\phi,
\label{eq:q_caust_j1}\qquad
y=\frac{4\beta_2}{\alpha}\sin^3\phi
\qquad \Rightarrow\qquad
\rho_c=\frac{4\beta_2}{\alpha}\Big(1-3\sin^2\phi\cos^2\phi\Big)^\frac{1}{2},
\label{eq:q_caust_j2}
\end{eqnarray}
where $\rho_c=\sqrt{x^2+y^2}$ is the radius vector of the caustic in the polar coordinate system $(\rho,\phi)$.

For a given target, the size of the quadrupole caustic of the SGL is set by the angle $\beta_s$ and the heliocentric distance to the observer's location in the image plane. For $\beta_s=8.13^\circ$ used in our simulations, at $r=650$~AU and $\lambda=1000$nm, the parameters $\alpha$ and $\beta_2$ are determined to be $\alpha\sim 48.97~{\rm m}^{-1}$, $\beta\sim 70.37$. The cusps of the caustic appear at four locations $\phi=\{0,{\textstyle\frac{\pi}{2}},\pi, {\textstyle\frac{3\pi}{2}}\}$ being placed at the distance of $\rho_c^{\tt cusp}=4\beta_2/\alpha\sim 5.75$~m. Similarly, the folds are at $\phi=\{{\textstyle\frac{\pi}{4}},{\textstyle\frac{3\pi}{4}}, {\textstyle\frac{5\pi}{4}},{\textstyle\frac{7\pi}{4}}\}$, at the distance of $\rho_c^{\tt fold}=2\beta_2/\alpha=2.87$~m.

These parameters yield an astroid caustic that is approximately 11.5 meters along a cusp-to-cusp direction and is 5.75 meters along a fold-to-fold diagonal. The images shown in this section are all simulations of a $20\times 20$ meter region in the image plane, which comfortably fits this astroid and the region immediately outside.

A telescope positioned at the center of the astroid caustic, $\rho=0$, of a point source of light ``sees'' a well-known pattern: four symmetrically positioned spots of light, the famous Einstein-cross, shown in Fig.~\ref{fig:psf} (right). A telescope that is positioned away from the optical axis, $\rho\not=0$, associated with a point source no longer sees such a symmetric pattern. Instead, a rich behavior emerges involving multiple spots and arcs of light, markedly different when the telescope is positioned inside vs. outside the astroid caustic. This is the behavior that we study in the present section, using a point source emitting monochromatic light at $\lambda=1~\mu$m. Our model telescope is a thin lens telescope with a $f=10$~m focal distance, projecting light onto an image sensor that is $200\times 200~\mu$m in size.

\begin{figure}
\includegraphics{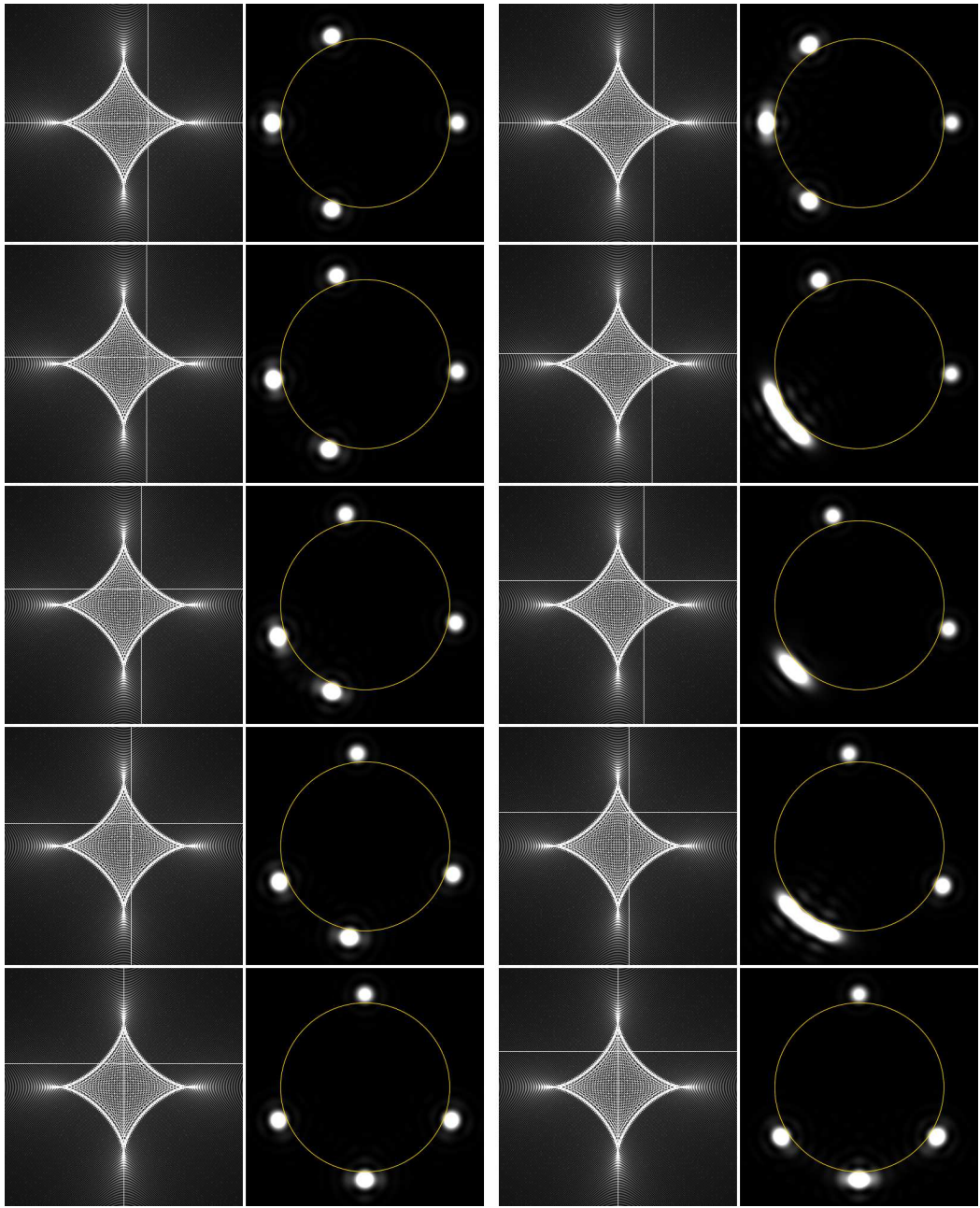}
\caption{\label{fig:inside}Left: The optical telescope moves around inside the caustic boundary, at a distance of $\rho=2$~m from the optical axis. First column: crosshairs mark the telescope position with respect to the caustic background; Second column: the view of the Einstein cross as seen by the telescope from the marked position. Right: As on the left, still mostly inside but closer to the boundary, at $\rho=3$~m. As the telescope approaches the caustic boundary, the two spots of lights that it sees merge into a partial arc. Note that the telescopic view is inverted with respect to the projected caustic: The caustic is viewed {\em from} the direction of the Sun, the telescope looks {\em towards} the direction of the Sun.}
\end{figure}

In the subsections below, we offer details and insight using key reference frames, which are taken from more lengthy animations, mini-``movies'' that show images that a telescope in motion might see. The full videos that correspond to these reference frames are available at \url{https://www.vttoth.com/CMS/physics-notes/355} as supplementary material to the present paper.

\subsection{Inside the astroid caustic}

Displacing the telescope from the center of the astroid caustic, we now study the resulting behavior. Fig.~\ref{fig:inside}, top left shows what happens to the view of the telescope if it is displaced by $\rho=2$~m from the optical axis in the image plane (i.e., away from the center of the caustics, but not yet at the fold $\rho<\rho_c^{\tt fold}$), as shown by the thin crosshairs in the image. The corresponding view in the telescope is a distorted version of the Einstein cross. In the image frames that are presented in the first two vertical columns of Fig.~\ref{fig:inside}, the telescope then begins to move in a circular pattern, completing a $90^\circ$ arc around the optical axis while maintaining a 2~m radial distance. As we can observe, the four spots of light that constitute the Einstein cross move about, their relative positions changing, but they radial distance from the center of the image sensor is unchanged. This corresponds to the fact that the impact parameter $b$ is determined by the solar monopole, so any light hitting the telescope aperture at a given distance from the Sun will always arrive from the same apparent distance from the solar limb.

\begin{figure}
\includegraphics{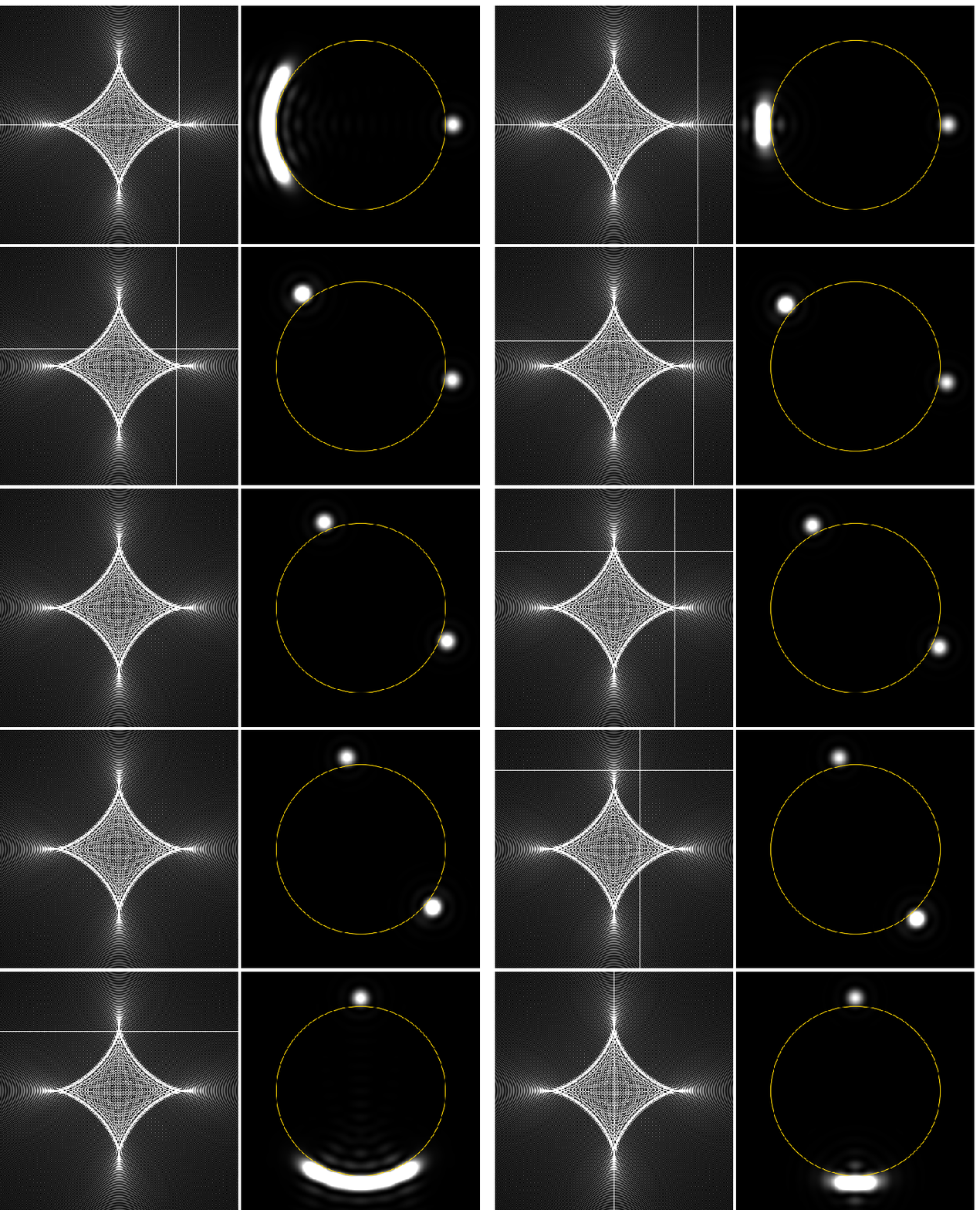}
\caption{\label{fig:outside}Left: The optical telescope is now mostly outside the caustic boundary, only encountering it briefly at the cusps, at a distance of $\rho=5$~m from the optical axis. First column: the telescope position with respect to the caustic, second column: the view of the Einstein cross as seen by the telescope. Right: As on the left, still further outside at $\rho=7$~m. No Einstein-cross is visible in these images; even as the telescope moves through the cusp region and sees a considerably widened arc, the arc does not split into multiple spots of light.}
\end{figure}

The rightmost two columns of Fig.~\ref{fig:inside} show the same view, except that in this case, the imaging telescope is now located at $\rho=3$~m from the center, which is $\rho\simeq \rho_c^{\tt fold}$. This distance is sufficient for the telescope to reach the position of the caustic boundary in the ``fold'' region of the astroid caustic, halfway between cusps. This has a profound impact on the image produced by the imaging telescope. When the telescope is in this region near the fold, two of the light spots from the Einstein cross merge, form an elongated, brightened arc of light. The arc is at its brightest when the telescope is closest to the caustic boundary. We can also see (middle row) that the arc begins to diminish rapidly as soon as the telescope is positioned outside the caustic boundary.

\subsection{Outside the astroid caustic}

Positioning the imaging telescope at a radial distance of $\rho=5$~m changes its view dramatically, as now we are in the vicinity of the cusp, $\rho\simeq\rho_c^{\tt cusp}$. At this radial distance and in the direction of one of the cusps, the telescope intersects the brightest region of the cusp of the caustic boundary, but elsewhere, it is now positioned firmly outside the boundary. At the cusp, as we can see in Fig.~\ref{fig:outside} (top left), three of the spots from the Einstein cross now merge into a bright, significantly elongated partial arc; only one standalone spot remains on the opposite side. As the telescope begins its angular motion, however, the arc rapidly collapse to a single spot of light; the other two spots are absent when the telescope is outside the caustic boundary. The bright arc reappears as the telescope completes a $90^\circ$ arc along its path, approaching the next cusp of the caustic boundary.

Finally, when we place the telescope even further outside the caustic boundary, $\rho=7~{\rm m}>\rho_c^{\tt cusp}$, the resulting image reduces to two spots, one of which brightens and elongates but only a little when the telescope is near the cusp regions. We can already discern in this behavior the emergence of the monopole pattern: a little further out, we recover the behavior seen in Fig.~\ref{fig:approach} (left), two identical spots of light at at the same distance from the Sun, on opposite sides. Thus we can see that outside the caustic boundary, the influence of the quadrupole moment diminishes rapidly.

\subsection{Light amplification}

\begin{figure}
\includegraphics{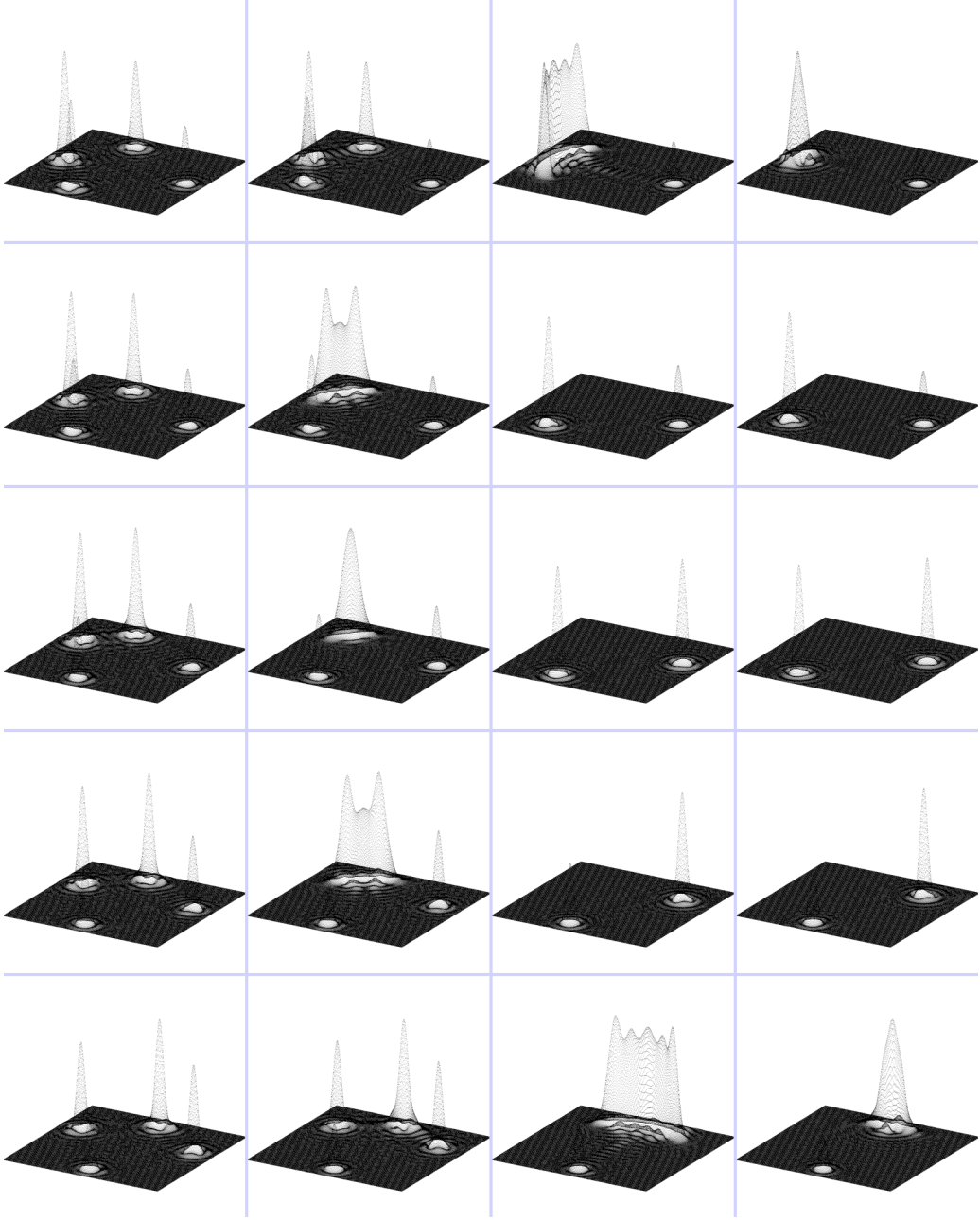}
\caption{\label{fig:amplitude}The intensity of light that is seen by an imaging telescope, corresponding to Figs.~\ref{fig:inside} and \ref{fig:outside}. The four columns correspond to telescope locations at 2, 3, 5 and 7~m from the optical axis, respectively. The five rows show five phases of the telescope's path as it traverses $90^\circ$ in the image plane while maintaining a constant distance frmo the optical axis. }
\end{figure}

Images such as Figs.~\ref{fig:inside} and \ref{fig:outside} offer a visual representation of the view seen by an imaging telescope, but cannot be readily used to judge the amount of light collected by such a telescope. For this reason, we prepared another series of plots that show the intensity of light on the telescope sensor in a three-dimensional form, on a third axis (Fig.~\ref{fig:amplitude}). The vertical scale is, of course, arbitrary (hence unlabeled) since it is determined by the intensity of the point source. In this figure, all four cases are shown (2~m through 7~m, in four columns) along with the phases of telescope motion that were presented in Figs.~\ref{fig:inside} and \ref{fig:outside}. As we can observe, the peak brightness between the images varies relatively little. However, when the telescope is positioned on the caustic boundary, especially near the cusps, it collects substantially more light due to the widening of the arcs that appear on its sensor. This is especially evident in the top and bottom images of the third column, corresponding to the telescope located at 5~m from the optical axis, corresponding to the approximate distance of the peak cusp region.

\subsection{Dependence on wavelength}

\begin{figure}
\includegraphics{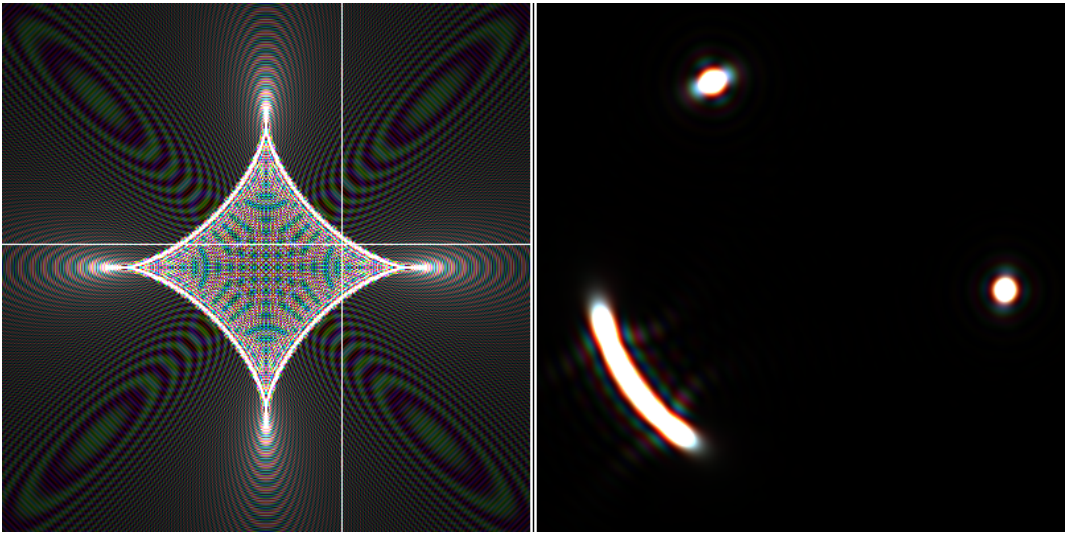}
\caption{\label{fig:color}The formalism that we introduced is entirely wave-optical in nature, and as such, makes it possible to model accurately color, including multispectral imagery. This is a representative image showing the SGL PSF and the view through an imaging telescope in the three RGB color channels that correspond to human color vision. Note that the caustic boundary is achromatic.}
\end{figure}

Finally, we must again emphasize that this analysis and the modeling tools that we developed take the wave-optical nature of light fully into consideration. As such, they offer the capability to model light at any wavelength, thus modeling observations using visible colors or other, multispectral observation.

By way of example, we present a snapshot view by the optical telescope at a distance of 3~m from the optical axis, but this time viewing light in the three wavelengths that characterize human vision: red (700~nm), green (530~nm) and blue (470~nm). Studying the color caustic in Fig.~\ref{fig:color} (left), we note that the overall geometry of the caustic boundary is unchanged. Moreover, the boundary itself is achromatic: light in the boundary region is predominantly white, indicating that the effect of the SGL PSF in these regions is independent of wavelength. Inside the caustic boundary, however, the interference pattern shows strong wavelength dependence: as a result, the colors of the rainbow appear. Similarly, outside the caustic boundary, the concentric pattern of the monopole PSF begins to emerge, but its spatial wavelength, too, depends on the wavelength of light; therefore, colors are separated and the colors of the rainbow emerge.

The telescopic view of the Einstein cross reflects these observations. The bulk of the light is white, albeit with a reddish tint: this tint emerges because of the dependence of the $J_1$ Bessel-function term in Eq.~(\ref{eq:Pvdd}), representing the thin lens telescope, on the wavenumber $k$. Only near the fringes of the spots or arcs of light seen by the telescope do we note the appearance of the colors of the rainbow, in the form of rapidly diminishing diffraction artifacts.

In this section, we presented a qualitative description of numerical results from modeling the PSF of the extended Sun and its convolution with a thin lens imaging telescope. These results will be used subsequently for numerical modeling of the imaging of extended objects such as an exoplanet, image convolution and deconvolution, and estimates of light amplification and noise. These efforts are presently ongoing.

\section{Discussion and Conclusions}
\label{sec:end}

We continue to investigate the gravitational deflection of light by extended gravitational lenses.  Based on the wave-optical method developed in \cite{Turyshev-Toth:2021-multipoles}, we treat the lens as an extended axisymmetric gravitating body that possesses an infinite set of zonal harmonics. Light propagation on such a background results in the formation of various caustics  \cite{Turyshev-Toth:2021-caustics}. Each multipole introduces a caustic of a particular shape (see Fig.~\ref{fig:caustics}) and,
in the case when several multipoles are present, an effective caustic is formed whose structure is now determined by the combined action of the multipoles.

In this paper, we investigated how the imaging of point sources  by the SGL is affected by the presence of the multipole caustics. For that, we used the known measured values of the leading zonal gravitational harmonics of the Sun. We have shown that the presence of the caustics affects the imaging  by modifying the structure of the imaging content projected on the Einstein ring. The ring is broken on several individual bright spots where the number of the spots is determined by the largest multipole, representing the symmetries relevant to that multipole.

In the case of the Sun, the astroid caustic caused by its quadrupole moment, $J_2$, dominates. The presence of other solar multipoles contribute only small perturbations to the astroid caustic of the quadrupole.

When viewed by an imaging telescope, the main consequence of the presence of the quadrupole moment is that the Einstein ring, which characterizes a monopole lens, is broken up into several spots and arcs. When the imaging telescope is located within the astroid caustic, the ring is broken in four bright spots, which are moving within the circumference of the ring on the focal plane of the imaging telescope as the telescope changes its position in the image plane. If the telescope is positioned exactly on the principal optical axis of the SGL, the spots of equal brightness are positioned exactly in the direction of the cusps of the caustic. However, as the telescope changes its position, both the relative brightness of the spots and their position in the ring change in a  fashion prescribed by (\ref{eq:B2}) and (\ref{eq:AinscER}).

When the telescope approaches one of the caustic cusps, a remarkable transformation takes place: three of the spots of light converge to form a partial arc. This arc eventually collapses into a single spot as the telescopic image transforms into the same image we see when viewing a point source through a monopole gravitational lens farther from the optical axis: two symmetrical spots of light (note that eventually these spots become asymmetric, but only at distances from the optical axis that are many orders of magnitude greater than what is being investigated here.)

When the telescope approaches one of the caustic folds, a different transformation can be seen: two of the spots merge into a single arc, which then rapidly disappears as the telescope moves outside the caustic region. The remaining two spots eventually transform into the image that is projected by a monopole SGL.

The image transformations discussed above are captured by the wave-optical treatment of the gravitational lensing phenomena \cite{Turyshev-Toth:2021-multipoles}. They relate the caustics formed on the image plane \cite{Turyshev-Toth:2021-caustics}, as given by (\ref{eq:B2}), to the images formed on the image sensor of an imaging telescope, as given by (\ref{eq:AinscER}). Understanding of these transformations is important for the on-going development of the deconvolution algorithms \cite{Toth-Turyshev:2020} that may be needed to recover high-resolution images of the faint sources.

These results and our investigation pave the way towards the imaging of extended, resolved objects by extended axisymmetric lenses, such as the SGL. Extended objects can, of course, be modeled as ensembles of point sources. The SGL PSF can be convolved with the source image to model light projected by the SGL in the image plane, or light seen by an imaging telescope. Inverting the process, that is, deconvolution, can be used to recover the original image from observational data.

Although our present focus is gravitational lensing by the Sun, our results are directly applicable to other astrophysical lenses that are characterized by axisymmetric gravitational fields and can be modeled using zonal harmonics.

These studies are ongoing. Results, when available, will be reported elsewhere.

\begin{acknowledgments}
This work in part was performed at the Jet Propulsion Laboratory, California Institute of Technology, under a contract with the National Aeronautics and Space Administration. VTT acknowledges the generous support of Plamen Vasilev and other Patreon patrons.

\end{acknowledgments}


\end{document}